\documentstyle[prl,aps,twocolumn]{revtex} \tolerance = 10000

\begin{document}


\title{Optimal Quantum Cloning Machines}

\author{N. Gisin$^{(1)}$ and S. Massar$^{(2)}$}
\address{(1) Group of Applied Physics, University of Geneva, 1211 Geneva,
Switzerland}
\address{(1) Raymond and Beverly Sackler Faculty of Exact Sciences,
School of Physics and Astronomy,
Tel-Aviv University, \\ Tel-Aviv 69978, Israel}

\date{\today}


\maketitle

\begin{abstract}
{\bf abstract} 
We present Quantum Cloning Machines (QCM) that transform N identical qubits
into $M>N$ identical copies and we
prove that the fidelity (quality) of these copies is optimal. 
The connection between cloning and measurement is discussed in detail.
When the number of clones M tends
towards infinity, the fidelity of each clone tends towards the
optimal fidelity that can be obtained by a measurement on the input
qubits. More generally, the QCM are universal devices to translate quantum
information into classical information.\\
\end{abstract}


Quantum Cloning Machines (QCM) act on an unknown quantum state and make
one, or more, copies of it. The superposition principle of quantum
mechanics prohibits the copies from being
perfect\cite{WZ}\cite{BCFJS}.
This basic result is maybe the most fundamental difference between
classical and quantum information theory, and QCM therefore probe in a
detailed way the structure of quantum information. 
For example, the $U_{1,2,}$ QCM 
makes two copies of one qubit (ie of a spin $\frac{1}{2}$ state) with a
fidelity independent of the
state of the input qubit \cite{Buzek}. Other recent related work has been
concerned with
deriving inequalities governing the quality of QCM\cite{Buzek2}, and
in applying QCM to concrete problems such as eavesdropping in quantum
cryptography\cite{GH}, quantum entanglement\cite{GH}\cite{BVPKH}\cite{Hor},
and building quantum computer networks to realize QCM\cite{BBHB}.

A conceptually simple cloning machine, which we shall call the
classical copying machine (CCM), is obtained by making a measurement
on the input state. The (classical) result of the measurement is then
used to make 
an arbitrary number $M$ of identical copies. Such a copying
machine only makes use of the  
information about the input state that
is available through measurement. 
It is therefore less efficient than the QCM's proper. 
Indeed the most general QCM consists of the $N$ input qubits all in the same
state, the $M-N$
blank copies all in the same neutral state, and an eventual ancilla, which
evolve unitarily into an (entangled) state of the $M$ clones and
the ancilla.
For any
finite number $M$ of copies, these QCM make better copies than the
CCM. But in the limit $M \to \infty$, the quality of the two copying
machines are equal. This shows that not only do QCM necessarily make
imperfect copies, but that information is necessarily diluted during the
copying process, since when $M$ is large the copies contain no more
information than that which is available classically. 
Note that the quantum information has not disappeared, but
is hidden in the correlations between the copies and the ancilla.

In this article, for simplicity, we concentrate on QCM that transform one qubit
into $M$ identical copies, though many results are also stated for an
arbitrary number
$N$ of input qubits. When $N=1$ and $M=2$ our QCM reduces 
to the QCM of Bu\v{z}ek 
and Hillery \cite{Buzek}. Furthermore we prove that these QCM are optimal,
i.e. that no other QCM can make better copies. 
We use this result to make precise the above discussion relating
cloning and measurement. 

The input state of our QCM is a qubit $|\psi> =
cos \theta/2 |\uparrow > + e^{i \phi} sin \theta/2 |\downarrow >$. The a priori
probability distribution
of the polarization direction $\psi$ is uniform over the Poincar\'e
sphere. We measure the quality of the copies by their fidelity ${\cal F}$,
ie. the mean overlap between any of the copies (they are all
identical) and the input state:
\begin{equation} 
{\cal F}=\int d\Omega <\psi |\rho_{out}|\psi >
\label{fud}\end{equation}
 where $\int d\Omega = \int_0^{2\pi} d\phi \int_0^\pi d\theta sin
\theta /4 \pi $, and $\rho_{out}$ is the reduced density matrix of 
one of the copies.

The $U_{1,M}$ QCM is described by the following unitary operator:
\begin{eqnarray}
U_{1,M}|\uparrow>\otimes R &=&
\sum_{j=0}^{M-1}\alpha_j|(M-j){}\uparrow,j{}\downarrow>\otimes R_j
\nonumber\\
U_{1,M}|\downarrow>\otimes R &=&
\sum_{j=0}^{M\!-\!1}
\alpha_{M\!-\!1\!-\!j}|(M\!\!-\!\!1\!\!-\!\!j)\uparrow,(j\!+\!1)
\downarrow>\otimes
R_j \nonumber\\
\alpha_j&=&\sqrt{\frac{2(M-j)}{M(M+1)}}
\label{Udown}
\end{eqnarray}
where $R$ denotes the initial state of the copy machine and the $M-1$
blank copies, 
$R_j$ are orthogonal normalized internal states of the
QCM, 
and we have denoted $|M-j
\psi, j \psi^{\perp}>$ the symmetric and normalized state with $M-j$
qubits in the state $\psi$ and $j$ qubits in the orthogonal state
$\psi^{\perp}$. 

A somewhat lengthy computation involving combinatorial series shows that
this unitary 
operator acts on an arbitrary input state $\psi$ as follows:
\begin{equation}
U_{1,M}|\psi>\otimes R =
\sum_{j=0}^{M-1}\alpha_j|(M-j){}\psi,j{}\psi^\perp>\otimes R_j(\psi)
\label{U1toM}
\end{equation}
where $R_j(\psi)$ represents the internal state
of our QCM with $R_j(\psi)\perp R_k(\psi)$ for all $j\ne k$.
In order to give a synthetic expression for $R_j(\psi)$,
let us introduce the qubits 
$\psi^* = cos \theta /2 |\uparrow^*> + e^{-i \phi} sin \theta/2  
|\downarrow^*>$
which transforms under rotations
as the complex conjugate representation. 
If we formally identify the internal states of the
QCM $R_j$ with the states 
$R_j = |(M-1-j){}\uparrow^*,j{}\downarrow^{*}>$, then the states
$R_j(\psi)$ are succinctly expressed as
$
R_j(\psi)=|(M-1-j){}\psi^*,j{}(\psi^
*)^\perp>$.

The density matrix describing the output qubits is the same for all
copies, and has
the form 
$\rho_{out} = {\cal F} | \psi> < \psi| + (1- {\cal F}) |\psi^\perp>
<\psi^\perp|$.
To calculate the fidelity ${\cal F}$ we first note that 
$\alpha_j^2$ is the probability that there
are $j$ errors among the M output
copies. 
Then, concentrating  on the first output
qubit, we have
\begin{eqnarray}
{\cal F}_{1,M}&=&\sum_{j=0}^{M-1}\tt{Prob(j\, errors\, in\, the\, M-1\, last\,
qubits)} \nonumber  \\
&=&\sum_{j=0}^{M-1}\frac{M-j}{M}\alpha_j^2=\frac{2M+1}{3M}
\label{fidel}
\end{eqnarray}
where $\frac{M-j}{M}$ is the ratio of the number of ways to chose $j$ errors
among $M-1$
qubits over the number of ways to chose $j$ errors among M qubits.
Note that since the possible final states of the QCM are orthogonal, one
can know whether
the copy process went through without error or not. However this requires a
priori knowledge
of the initial state, since the two possible final states of the QCM depend
on it. If one does not have any a priori knowledge about the initial
state (and this is what we assume) then it is 
impossible to learn by making a measurement on the 
QCM whether or not the cloning has succeeded.

We have also constructed a more general QCM that takes $N$  
identical input qubit into  $M$ identical copies. It is described by
\begin{eqnarray}
U_{N,M}|N{}\psi> &=& \sum_{j=0}^{M-N}\alpha_j
|(M-j){}\psi,j{}\psi^\perp>\otimes R_j(\psi)
\nonumber\\
\alpha_j &=&
\sqrt{\frac{N+1}{M+1}}\sqrt{\frac{(M-N)!(M-j)!}{(M-N-j)!M!}}
\label{UNtoM}
\end{eqnarray}
where $|N{}\psi>$ is the input state consisting of $N$ spins all
in the state $\psi$, and the other notations are as above.
Note that the number $j$ of errors in the copies is smaller or equal to the
number $M-N$ of additional qubits.
The fidelity of each output qubit is
\begin{eqnarray}
{\cal F}_{N,M}
&=&\sum_{j=0}^{M-1}\frac{M-j}{M}\alpha_j^2=\frac{M(N+1) +N}{M(N+2)}
\end{eqnarray}
The N to N+1 cloning machine is particularly simple since
the right hand side in eq. (\ref{UNtoM}) contains only two terms. In this
case the fidelity ${\cal F}_{N,N+1} = \frac{N^2 + 3N +1 }{N^2 + 3N +
  2}$ tends rapidly towards 1 as N grows, corresponding to the fact
that the input state is quasi--classical.

The fidelity ${\cal F}_{N,M}$ of these QCM (\ref{UNtoM}) tends to $N+1/N+2$ for
large $M$
which is the optimal
fidelity achievable by carrying out a measurement on $N$ identical
qubits
\cite{MassarPopescu1995}. This suggests that the QCM tends towards the
CCM as $M$ increases. We now prove that this is indeed the case. Let
us first consider the case $N=1$.
In \cite{MassarPopescu1995} it was shown that an optimal measurement
on a single qubit is simply a Stern Gerlach measurement, that is a
projection onto two (randomly chosen) orthogonal states
$|\phi>$ and $|\phi^\perp>$. The corresponding CCM consists of making 
$M$ copies of the $\phi$ state ($|M {}\phi>$) 
if the outcome of the measurement 
is $\phi$,
and $M$ copies of the $\phi^\perp$ state ($|M {}\phi^\perp>$) 
if the outcome of the measurement 
is $\phi^\perp$.
The density matrix describing the
$M$ copies, averaged over the orientations of the measuring basis
$|\phi>$, 
is 
\begin{eqnarray}
\rho_{CCM} &=& \int d \Omega_{\phi} 
|<\psi|\phi>|^2 P_{|M{}\phi>}\nonumber\\ & & 
+ |<\psi|\phi^\perp>|^2
P_{|M{}\phi^\perp>}
\end{eqnarray}
where the first factor is the probability to have outcome $\phi$ 
($\phi^\perp$), and
$P_{|M{}\phi>}$ ($P_{|M{}\phi^\perp>}$)
is the projector onto the state
$|M{}\phi>$ ($|M{}\phi^\perp>$). 
In order to compare the CCM to the QCM, we express
$\rho_{CCM}$ in the basis $\psi$, $\psi^\perp$ to obtain:
\begin{equation}
\rho_{CCM} =
\sum_{s=0}^{M}\frac{2 (M+1 -s)}{(M+1)(M+2)}
P_{|(M\!-\!s){}\psi,s{}\psi^\perp>}
\end{equation}
It is then easy to show that the QCM tends towards the CCM as
$M$ increases. For instance one has 
$Tr [\rho_{QCM} -  \rho_{CCM}]^2 
\simeq M^{-3}$. Other measures of the ``distance''
between $\rho_{CCM}$ and $\rho_{QCM}$ similarly decrease 
as $M$ increases. A more complicated procedure, based on the
measurement eq. (15) of \cite{MassarPopescu1995}, shows that 
for an arbitrary number $N$ of input qubits
\begin{eqnarray}
\rho_{CCM} =
(N\!+\!1)\sum_{s=0}^{M}\frac{M!(M\!+\!N\!-\!s)!}{(M\!+\!N\!-\!s)!(M\!-\!s)!}
P_{|(M\!-\!s){}\psi,s{}\psi^\perp>}  \nonumber
\end{eqnarray}
and $Tr [\rho_{QCM} -  \rho_{CCM}]^2 \simeq N^4M^{-3}$. 
Thus when the number of copies $M$ increases, the QCM tends
towards the CCM. 
Conversely 
one can consider QCM as measuring
devices. Indeed, 
given N qubits all in the same unknown state $\psi$, one can
either make
a coherent measurement of all N qubits, or equivalently use the QCM to
produce a very large
number $M$ of clones and then do separable measurements on the clones. 
Indeed, since for large $M$, 
$\rho_{QCM}$ is a mixture of product states of the
form $|M\times\phi>$ it suffices to measure them with a classical polarimeter.
The fidelity of these two ways of gaining information about $\psi$ are
equal.  Hence the QCM can be considered as
a
universal device
transforming quantum information into classical information.
This is illustrated in figure 1. 
Note that in
the first case, all the difficulty for experiments lies in the coherent 
measurement, whereas in the second case
all the difficulty is in the QCM.

We now prove that the QCM we have described are optimal.
For simplicity we 
consider the case where their is only one input qubit, but an
arbitrary number $M$ of output qubits. 
The idea of the calculation follows closely the analysis of optimal
measurements of \cite{MassarPopescu1995}. We first express in full
generality the average
fidelity ${\cal F}$ of a quantum cloning machine in terms of the
final state $|R_{jk}>$ (see below) of the machine. These final states
are subject to the condition that the evolution is unitary. One must
then maximize ${\cal F}$ subject to the unitarity conditions which are
introduced by using Lagrange
multipliers. The problem then reduces to an eigenvalue equation
for a matrix $A$, and the extremal value of ${\cal F}$
is expressed in terms of the largest eigenvalue of this matrix.

The most general QCM acts on the input qubits $\uparrow$, $\downarrow$ 
in the following way
\begin{equation}
|j>|R> \to |M-k \uparrow, k \downarrow >|R_{jk}> \quad j = \uparrow , 
\downarrow
\end{equation}
where $|R>$ is the initial state of the QCM and the blank copies,
$|R_{jk}>$ are unnormalized final states of the ancilla,
and we use a summation convention: repeated indices are summed over.
Unitarity of the evolution imposes that
\begin{equation}
 <R_{j'k}|R_{jk}>= \delta_{j' , j}
\label{unit}
\end{equation}
Note that because $|M-k \uparrow, k \downarrow >$ is 
completely symmetric, we have
made the hypothesis that the output of the QCM is completely
symmetric. As discussed below, 
this hypothesis can be dropped without affecting
our conclusions.
Our task is to maximize the fidelity of this QCM subject to the
unitary constraints (\ref{unit}). The
rotational symmetry of the input qubits is exploited by expressing an
arbitrary input qubit as a SU(2) rotation $O_{j'j}(\Omega)$
acting on the $\uparrow$ state:
$|\psi> = cos \theta/2 \uparrow + e^{i \phi} sin \theta /2 \downarrow =
O_{\uparrow j} |j>$. The
evolution of an arbitrary input qubit is then
\begin{eqnarray}
&|\psi> |R> = O_{\uparrow j} |j> |R>& \nonumber\\
&\to |\psi_{out}>= O_{\uparrow j} |M-k \uparrow, k \downarrow>|R_{jk}>& 
\end{eqnarray}
Because the output state is symmetric under permutations, the fidelity
of the copies is obtained by calculating the overlap of the reduced
density matrix of one copy, say the first, with the input state $|\psi>$
and averaging over the input states.
One finds
\begin{eqnarray}
{\cal F} &=& Tr [ < \psi_{out} |
O_{\uparrow i'} |i'><i|O^{*}_{\uparrow i} | \psi_{out}> ]\quad 
i,i'=\uparrow,\downarrow
\nonumber\\
&=& <R_{j'k'}|R_{j k}>
\left(\int d\Omega O^{*}_{\uparrow j'}
O_{\uparrow i'} O^{*}_{\uparrow i} O_{\uparrow j}\right)\nonumber\\
& & \quad {} Tr \left[<M-k' \uparrow,k'\downarrow| |i'>< i | 
|M-k\uparrow ,k \downarrow>
\right]\nonumber\\
&=&<R_{j'k'}|R_{j k}>A_{j',k',j,k}
\label{expF}
\end{eqnarray}
where we have expressed everything in terms of the SU(2) rotation
matrices and introduced the matrix $A_{j',k',j,k}$ which plays an
essential role in this calculation. Our problem is to maximize ${\cal F}$
subject to the unitary constraints eq. (\ref{unit}).
We impose the unitary constraints by adding them via 
Lagrange multipliers $\lambda_{j'j}$.  Thus we must extremize
\begin{eqnarray}
{\cal F}&=&<R_{j'k'}|R_{j k}>A_{j',k',j,k}
\nonumber\\
& &-\lambda_{j'j}\left(<R_{j'k'}|R_{jk}>\delta_{k',k}- \delta_{j' ,
j}
\right)
\label{fullP}\end{eqnarray}
with respect to the final states of the QCM $|R_{j k}>$ and the
multipliers $\lambda_{j'j}$. It is however useful to consider a
simpler problem in which we impose only one constraint, namely the
trace of eq. (\ref{unit}).  Obviously the extrema of this reduced
problem are greater or equal to the extrema of the full problem
eq. (\ref{fullP}), and we will thus obtain an upper bound on the
fidelity of QCM. We shall show below that rotational symmetry
implies that this upper bound is attained by optimal QCM. 
Thus we have to extremize
\begin{eqnarray}
{\cal F}&=&<R_{j'k'}|R_{j k}>A_{j',k',j,k}\nonumber\\
& &
-\lambda\left(<R_{j'k'}|R_{jk}>\delta_{k',k}\delta_{j',j}- 2
\right)
\label{P}\end{eqnarray}
Varying with
respect to $<R_{j'k'}|$ (more properly one should vary with respect to  
the components of $<R_{j'k'}|$ in a basis), we obtain the equations
\begin{equation}
\left(
A_{j',k',j,k} -\lambda\delta_{k',k}\delta_{j',j}\right)|R_{jk}> =0
\label{EQ}
\end{equation}
Thus $
\lambda$ are the eigenvalues of $A_{j',k',j,k}$ and
$|R_{jk}>$ its eigenvectors.
Suppose we have found a solution $\lambda , |R_{jk}>$ of eq.
(\ref{EQ}) and of the unitary constraints eq. (\ref{unit}). 
Then multiplying eq. (\ref{EQ})
on the left by $ <R_{j'k'}|$ and summing over $j',k'$ yields
\begin{eqnarray}
&<R_{j'k'}|R_{jk}>A_{j',k',j,k}& \nonumber\\ & = 
<R_{j'k'}|R_{jk}>\lambda\delta_{k',k}\delta_{j',j} = 2 \lambda &
\end{eqnarray}
where the last equality follows from the constraint eq. (\ref{unit}).
But the left hand side is equal to the fidelity ${\cal F}$
eq. (\ref{expF}). 
So the eigenvalues $\lambda$ of $A$ are related to the optimal
fidelity of the QCM by
$
{\cal F} = 2 \lambda 
$.
It remains to calculate the matrix $A$. After some algebra one finds
that it is block diagonal $A_{ j',k',j,k} =\delta_{k-j,k'-j'}B_{k+j,k'
+j'}$ with 
\begin{equation}
B = {1 \over 6 M}\left ( 
\begin{array}{cc}
{2 M - K }&{\sqrt{(M-K)(K+1)}}\\
{\sqrt{(M-K)(K+1)}}&{M+K+1}
\end{array}\right)
\end{equation}
where $K = k-j$.
The largest eigenvalue of $A$ is $(2M +1) /6M$ corresponding to an
upper bound on the
optimal fidelity ${\cal F} \leq (2M +1)/3M$.
This bound is saturated by the QCM eq. (\ref{Udown}) thereby proving
that it is optimal.
We have generalized this proof to show that the
QCM eq. (\ref{UNtoM}) that transform $N$ identical
qubits   into an arbitrary number $M$ of copies are optimal. Our proof
is at present only valid for $N=1,2,...,7$, although we expect it to
generalize to arbitrary $N$. The difficulty when
$N$ is large is that the matrix $B$ is $N+1\times N+1$ and its
eigenvalues are correspondingly difficult to calculate. 

Throughout this letter we have considered QCM that are symmetric in
their output qubits. If we want to generalize our results to QCM that
are not symmetric, we must adopt a more general definition of the
fidelity of the copies which we take to be the average fidelity of
each copy. We will now show that 
there necessarily exist QCM that are optimal in this more general
sense and symmetric. Indeed, suppose one has
constructed a (not necessarily symmetric) optimal QCM. One can then
build another QCM which is identical to the proceeding one,
except that some of the output states have been permuted. This QCM is
obviously also optimal because of the symmetry of the definition of fidelity.
If one takes a coherent superposition of these QCM, averaged over all
possible permutations of the output qubits, one obtains still another
optimal QCM, but which is symmetric in its output qubits.

One can similarly show that their necessarily exist optimal QCM that
are rotation invariant. Indeed suppose that one has built a (not
necessarily 
rotationally invariant) optimal QCM. By rotating the whole apparatus one
obtains another optimal QCM. (This follows from the rotation
invariance of the definition of fidelity eq. (\ref{fud}), which is
averaged over all possible orientations of the input qubits). If one
takes a coherent superposition of these QCM, averaged over the
orientations of the apparatus, one obtains an optimal and 
rotationally invariant
QCM \cite{optQC}. We further note that the Lagrange 
multipliers $\lambda_{jj'}$
associated to such an optimal rotationally invariant QCM must also be
invariant under rotation, which can only be the case if
$\lambda_{jj'}=\lambda
\delta_{jj'}$ (Shur's Lemma). This explains why the extrema of the
reduced problem eq. (\ref{fullP}) are also extrema of the full 
problem eq. (\ref{P}).

In summary, 
the QCM that have been presented are optimal as well for copying quantum
information
as for translating quantum information to classical information, thus
establishing the connection between cloning quantum information and 
gaining classical information. 
For example, the $1\rightarrow 2$
QCM provides the optimal eavesdropping strategy for a quantum cryptography
protocol based
on 3 non-orthogonal bases X, Y and Z on the Poincar\'e sphere, as
conjectured by C. Fuchs [private communication]. 
The experimental realization
of such optimal QCM is a worthwhile challenge. Indeed, it would provide a
universal device
for copying and reading quantum information.\\

We would like to thank V. Buzek, A. Ekert, L. Goldenberg, 
B. Huttner, Ch. Macchiavello, S. Popescu and D. Rohrlich for helpful
discussions. 
NG acknowledges financial support by the Swiss National
Science Foundation and by the European TMR network
ERB--FMRX--CT96--0087 and SM acknowledges support from grant 614/95 of
the Israel Science Foundation.


\section*{Figure Captions}

Fig. 1:  Diagram of the flow of quantum information to classical information.


\begin{thebibliography}{99}

\bibitem{WZ} W. K. Wootters and W. H. Zurek, Nature {\bf 299}, 802 (1982)

\bibitem{BCFJS} H. Barnum, C. M. Caves, C. A. Fuchs, R. Jozsa and
  B. Schumacher, Phys. Rev. Lett. {\bf 76} (1996) 2818

\bibitem{Buzek}
V. Bu\v{z}ek and M. Hillery, Phys. Rev. A {\bf 54}, 1844 (1996).

\bibitem{Buzek2}
 M. Hillery and V. Bu\v{z}ek,  {\it Quantum Copying: Fundamental
 Inequalities}, quant-ph/9701034

\bibitem{GH} N. Gisin and B. Huttner, {\it Quantum Cloning,
    Eavesdropping and Bell's inequality}, Phys. Lett. A {\bf 228}, 13, 1997.

\bibitem{BVPKH} V. Bu\v{z}ek, V. Vedral, M. B. Plenio, P. L. Knight
  and M. Hillery, {\it Broadcasting of entanglement via local
  copying}, quant-ph/9701028

\bibitem{Hor} M. Horodecki and R. Horodecki, {\it Is their a basic law
of quantum information processing}, quant-ph/9705003


\bibitem{BBHB} V. Bu\v{z}ek, S. L. Braunstein, M. Hillery and D. BruB,
{\it Quantum copying: A network}, quant-ph/9703046

\bibitem{MassarPopescu1995} S. Massar and S. Popescu, Phys. Rev. Lett. {\bf
74}, 1259, 1995.

\bibitem{optQC} 
A similar argument has been used to discuss the symmetry of optimal
measurements: A. S. Holevo, Journal of Multivariate Analysis {\bf 3} (1973)
337; and the symmetry of eavesdropping
strategies in quantum cryptography: I. Cirac and N. Gisin, 
{\it Coherent eavesdropping strategies for the 4 state 
quantum cryptography protocol}, Phys. Lett. A, in press, 1997, quant-ph 9702002;
Ch. Fuchs et al. {\it Optimal eavesdropping in quantum cryptography}, Phys.
Rev. A, 
in press, 1997, quant-ph 9701039.


\end{thebibliography}
\end{document}